\documentstyle[aps,multicol,epsfig]{revtex}

\newcommand{\simle}
{\raisebox{-0.75ex}[-1.5ex]{$\;\stackrel{<}{\sim}\;$}}
\newcommand{\simge}
{\raisebox{-0.75ex}[-1.5ex]{$\;\stackrel{>}{\sim}\;$}}
\def\d{{\partial}}
\def\s{{\sigma}}
\def\e{{\epsilon}}
\def\k{{ {\bf k} }}

\def\w{{\omega}}

\begin{document}

\def\runtitle{
Generalized Kadowaki-Woods Relation 
in Heavy Fermion Systems with Orbital Degeneracy
}
\def\runauthor
 {Hiroshi {\sc Kontani}}

\title{
Generalized Kadowaki-Woods Relation \\
in Heavy Fermion Systems with Orbital Degeneracy
}

\author{
Hiroshi {\sc Kontani}
}

\address{
Department of Physics, Saitama University,
255 Shimo-Okubo, Saitama-city, 338-8570, Japan.
}

\date{\today} 

\sloppy
\maketitle

\begin{abstract}
{
We present a theoretical study of the Kadowaki-Woods relation 
in the orbitally degenerate periodic Anderson model.
Based on Fermi liquid theory,
we derive {\it the generalized Kadowaki-Woods relation
in the strong coupling limit},
$A\gamma^{-2} \approx 10^{-5} /\frac12 N(N-1)$ 
[$\mu\Omega$cm(mol$\cdot$K/mJ)$^2$],
where $A$ is the coefficient of the $T^2$ term in the resistivity,
$\gamma$ is the $T$-linear specific heat coefficient,
and $N$ is the $f$-orbital degeneracy.
This result naturally explains the 
remarkably smaller value of $A\gamma^{-2}$
in various orbitally degenerate (mainly Yb-based) heavy Fermion systems,
reported by Tsujii {\it et al.}, 
J. Phys. Cond. Mat. {\bf 15} (2003) 1993.
}
\end{abstract}

\vspace{5mm}

\vspace{5mm}

\begin{multicols}{2}

In Fermi liquid (FL) systems,
the specific heat $C$ and the resistivity $\rho$
behave as $C = \gamma T$ and $\rho= \rho_0+AT^2$
at sufficiently low temperatures.
Because $\gamma \propto m^\ast$ and $A\propto (m^\ast)^2$
($m^\ast$ being the effective mass of quasiparticles)
according to the FL theory, the ratio 
$A\gamma^{-2}$ is expected to be independent of $m^\ast$.
In fact, many Ce- and U-based heavy Fermion (HF) systems
follow a universal relation, 
$A\gamma^{-2} \approx 1\times10^{-5}$
[$\mu\Omega$cm(mol$\cdot$K/mJ)$^2$],
which is called the Kadowaki-Woods (KW) relation
 \cite{Kadowaki}.
Moreover, this relation also holds in 
$d$-electron heavy electron
systems like LiV$_2$O$_4$
($\gamma = 350 {\rm mJ/mole \cdot K}^2$)
and A-15 compounds
($\gamma \sim 15 {\rm mJ/mole \cdot K}^2$)
 \cite{Tsujii}.
Thus, the KW relation has been considered as
one of the remarkably robust signature
of Fermi liquids, irrespective of the value of $m^\ast$.
It can be derived microscopically 
using the FL theory 
 \cite{Yamada-PAM},
or using the slave-boson method
 \cite{slave}.
Note that the KW relation holds experimentally
even in the close vicinity of a magnetic quantum critical point 
under a magnetic field 
 \cite{Gegenwart},
the fact of which is consistent with theoretical analyses
 \cite{Takimoto,Continentino,Coleman}.

% Yb-compounds
Recently, however,
various Fermi liquid systems 
which does not follow the KW relation 
have been found experimentally.
Especially, 
Tsujii {\it et al.} have revealed that 
$A\gamma^{-2} \approx 0.4\times10^{-6}$
[$\mu\Omega$cm(mol$\cdot$K/mJ)$^2$]
in many Yb-based HF systems like YbCu$_4$Ag 
($\gamma = 240 {\rm mJ/mole \cdot K}^2$),
YbCu$_{5-x}$Ag$_x$ ($210\sim460 {\rm mJ/mole \cdot K}^2$), 
%YbCu$_{4.5}$, YbNi$_2$Ge$_2$, 
and others
 \cite{Tsujii,Tsujii2,Tsujii3}.
It is about 20$\sim$30 times smaller than the conventional 
KW ratio, although they are expected to be Fermi liquids.
Thus, the violation of the KW relation
should be a very important and fundamental subject 
on the FL theory.
The authors of ref.
 \cite{Tsujii}
suggest that materials with smaller $A\gamma^{-2}$
have {\it almost fully degenerate ground states}.
In fact, the crystalline electric field (CEF)
in Yb-based HF systems is in general smaller than that in 
Ce-based HF ones because Yb$^{3+}$-ion is smaller than 
Ce$^{3+}$-ion, known as lanthanoid contraction.
Note that a smaller value of $A\gamma^{-2}$ 
is also observed in Pd, Pt, Ni and Fe
where $\gamma \simge \gamma_{\rm band}$
 \cite{Miyake}.

% punch line
In the present work,
we revisit the KW relation in HF systems
based on the FL theory,
by taking the $f$-orbital degeneracy into account.
By applying the diagrammatic method
developed in analyzing the impurity Anderson model
%by several authors
 \cite{Yamada-AM,Shiba,Yoshimori},
we succeed in deriving a generalized KW relation,
eq. (\ref{eqn:Agamma}),
which is valid in the strong coupling case where 
$m^\ast \gg m_{\rm band}$.
By putting $N=8$ ($N$ being the orbital degeneracy)
which corresponds to a free Yb$^{3+}$-ion, 
we obtain that 
$A\gamma^{-2} \sim 0.36\times10^{-6}$
[$\mu\Omega$cm(mol$\cdot$K/mJ)$^2$],
which is consistent with experimental observations.
This is the first theoretical derivation of the KW relation
in orbitally degenerate systems.

% Miyake

As discussed in ref.
 \cite{Tsujii},
the violation of the KW relation
in Yb-compounds cannot be ascribed to the impurity effect
nor an accidental singularity of the band structure
inherent to individual compounds.
For this reason, we attack this universal issue 
based on a conventional orbitally degenerate
periodic Anderson model (OD-PAM) with a single conduction band.
Because of the strong $L$-$S$ coupling,
the $f$-electron state for Yb$^{3+}$-ion with $4f^{13}$
(Ce$^{+3}$-ion with $4f^{1}$)
is specified by the total angular momentum $J=7/2$ ($J=5/2$)
and its $z$-component $M$.
The degeneracy of $f$-orbital is $N=2J+1$.

Here, we study the following OD-PAM
 \cite{Hanzawa-PAM,Kontani-susc,Kontani-SOPA}:
\begin{eqnarray}
H &=& \sum_{\k}\e_{\k\s}c^\dagger_{\k\s}c_{\k\s} 
 + \sum_{\k M} E_{\rm f} f^\dagger_{\k M}f_{\k M}
 \nonumber \\
& & + \sum_{\k M\s} \left( 
 V_{\k M\s}^\ast f^\dagger_{\k M}c_{\k\s} + {\rm h.c.} \right)
 \nonumber \\
& & + \frac{U}{2}  \sum_{i M\neq M'}
 n_{i M}^f n_{i M'}^f ,
 \label{eqn:Ham}
\end{eqnarray}
where $c_{\k \sigma}^\dagger$ ($f_{\k M}^\dagger$)
is the creation operator of a $c$(conduction)-electron with
spin $\sigma=\pm \frac12$ ($f$-electron with $M=J,J-1,\cdots,-J$),
$n_{i M}^f = f_{i M}^\dagger f_{i M}$,
$\e_\k$ is the dispersion for $c$-electrons,
and $E_{\rm f}$ is the localized $f$-level energy.
Here we assume $E_{\rm f}>\mu$
($\mu$ being the chemical potential)
commonly for Ce and Yb-based HF systems,
by considering the particle-hole transformation for 
Yb$^{3+}$-ion.
The $c$-$f$ mixing potential matrices are  
given by $V_{\k M\s} = V_0\cdot\delta_{M,\sigma}$
for non-orbitally-degenerate case ($J=1/2, N=2$),
and it is given by
$V_{\k M\s}= V_0\sqrt{\pi} \sqrt{ (7/2+2M\s)/7 }
  Y_{l=3}^{M-\sigma} (\theta_k ,\varphi_k )$
for Yb$^{3+}$ ($J=7/2, N=8$), and
$V_{\k M\s}= 2\sigma V_0 
\sqrt{4\pi/3} \sqrt{ (7/2-2M\s)/7 } 
  Y_{l=3}^{M-\sigma} (\theta_k ,\varphi_k )$
for Ce$^{3+}$ ($J=5/2, N=6$),
where $Y_{l}^m(\theta_k ,\varphi_k )$ is the spherical harmonics.
We note that the relation 
$\sum_{M=-J}^{J} |V_{\rm k M\s}|^2 = V_0^2$
is satisfied in each case.

The $f$-Green function of the present model, $G_{\k MM'}(\e)$,
is given in ref. 
 \cite{Kontani-susc}.
It is shown that $G_{\k MM'}(\e)\propto e^{i(M-M')\varphi_k}$,
so it vanishes except for $M'=M$ after the summation taken over $\k$.
Here we study the three-dimensional OD-PAM 
within the framework of
the dynamical-mean-field-approximation (DMFA)
where the $d=\infty$ limit is taken systematically
 \cite{Vollhardt,Kotliar}.
In the DMFA, the self-energy $\Sigma(\e)$ is constructed of
local $f$-Green function(s),
$g(\w)\equiv \frac1{N_0} \sum_\k G_{\k MM}(\w)$,
which is given as
 \cite{Kontani-susc,Kontani-SOPA}
\begin{eqnarray}
& &g(\w)= \frac2{N} G^f(\w) 
 + \left( 1-\frac{2}{N} \right) G^{f0}(\w), 
 \label{eqn:green1} \\
& &G^f(\w)= \frac 1{N_0} \sum_k \left({1/G^{f0}(\w)- V_0^2/(\w+\mu-\e_\k)}
 \right)^{-1},  \label{eqn:green2} \\
& &G^{f0}(\w)= \left({\w+\mu-E_f-\Sigma(\w)} \right)^{-1}
 \label{eqn:green3} ,
\end{eqnarray}
where $G^{f0}$ represents the
$f$-Green function without mixing with $c$-electrons.
$N_0$ is the number of sites.
Note that $g(\w)$ is diagonal with respect to $M$
and independent of $M$. 
Below, we will utilize this fact to analyze
the strong coupling case.

First, we discuss the charge susceptibility $\chi_{\rm c}$
and the specific heat coefficient $\gamma$,
which are given as
\begin{eqnarray}
\chi_{\rm c} 
%&\equiv& \frac{\d n_M}{\d\mu} 
% \nonumber \\
&=& \frac1{z_\mu} N\rho^f(0)
%\left(1-\frac{\d\Sigma(0)}{\d\mu}\right) N\rho^f(0)
 \label{eqn:charge}, \\
\gamma &=& \frac{\pi^2}{6} \frac1z N\rho^f(0)
 \label{eqn:gamma},
\end{eqnarray}
at zero temperature, where 
$1/z\equiv 1-\d\Sigma(\e)/\d\e|_{\e=0}$,
$1/z_\mu \equiv 1-\d\Sigma(0)/\d\mu$, and 
$\rho^f(0)\equiv {\rm Im}g(-i\delta)/\pi$
is the $f$-electron density of states (DOS) 
per channel at the Fermi level;
the total $f$-electron DOS is $N\rho^f(0)$.

\begin{figure}
%\vspace{10mm}
\begin{center}
\epsfig{file=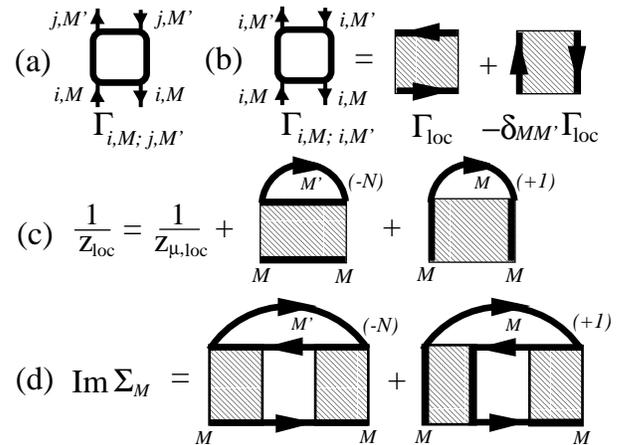,width=8cm}
\end{center}
%\vspace{10mm}
\caption{
(a) a full four-point vertex.
(b) a full local four-pont vertex, which is
given by $(1-\delta_{MM'})\Gamma_{\rm loc}$.
$\Gamma_{\rm loc}$ is an asymmetric local vertex.
(c) expression for $1/z_{\rm loc}-1$.
(d) expression for Im$\Sigma$.
In (c) and (d), the factor (-N)
originates from the summation over $M'$
and the Fermion loops.
}
  \label{fig:diagram}
\end{figure}

By shifting the frequencies of every closed loop 
in $\Sigma(\w)$ by $\w$, we obtain the identity
 \cite{Yamada-PAM,Yoshimori}
% \cite{Yamada-PAM,Yoshimori,Nozieres,AGD}
%
\begin{eqnarray}
%\gamma &=& \left.\left(1-\frac{\d\Sigma(\w)}{\d\w}\right)\right|_{\w=0}
% N\rho^f(0)  \nonumber \\
\frac{1}{z} 
 = \frac{1}{z_\mu}
%1-\frac{\d\Sigma(0)}{\d\mu} 
 + \sum_{j,M'} \Gamma_{i,M;j,M'}(0,0)\rho^f(0) ,
 \label{eqn:dsdm}
\end{eqnarray}
where 
$i,j$ are site indices, and 
$\Gamma_{i,M;j,M'}(\e,\e')$ is the full four-point vertex
within the DMFA; see Fig. \ref{fig:diagram}(a).
In the strong coupling limit where $U$ is sufficiently large,
$\gamma$ will be strongly enhanced whereas
$\chi_{\rm c}$ is suppressed.
This means that $1/z \gg 1/z_\mu$, so $1/z_\mu$
in eq.(\ref{eqn:dsdm}) can be dropped in the strong coupling case.

Here,
we introduce a modified renormalization factor
$1/z_{\rm loc}$ by dropping $\Gamma_{i,M;j,M'}$ 
except for $j=i$ in eq.(\ref{eqn:dsdm}), 
which we call the local approximation.
We note that by adding the term 
$U\sum_{i M} n_{i M}^f n_{i M}^f = U\sum_{i M} n_{i M}^f$
to eq.(\ref{eqn:Ham}) virtually,
we can neglect the Pauli principle
in the Coulomb interaction in constructing diagrams
 \cite{Yoshimori,Kontani-susc}.
This added term 
can be absorbed by shifting $E_f$ in eq.(\ref{eqn:Ham})
because it is a $M$-independent constant 
in the paramagnetic uniform state.
Then, the identity
$\Gamma_{i,M;i,M'}(0;0)= (1-\delta_{MM'})\Gamma_{\rm loc}(0,0)$ 
is easily recognized
within the DMFA as shown in Fig. \ref{fig:diagram}(b),
where $\Gamma_{\rm loc}(0,0)$ is the asymmetric local vertex
composed of local Green functions and $U$'s.
We can also check this identity order by order with respect to $U$
 \cite{Yoshimori}.
Taking notice of the factor 
arises from the summation over $M'$,
we find that
$1/z_{\rm loc}=(N-1)\Gamma_{\rm loc}(0,0) \rho^f(0)$;
see Fig. \ref{fig:diagram}(c).
As a result,
the specific heat coefficient in the local approximation,
$\gamma_{\rm loc}$, is given as
\begin{eqnarray}
\gamma_{\rm loc}
%&=& \frac{\pi^2 N_{\rm A}}{6 n} N(N-1)
&=& \frac{\pi^2}{6} N(N-1)
 \Gamma_{\rm loc}(0,0) \rho^f(0)^2
 \label{eqn:gamma_loc},
\end{eqnarray}
when $1/z_{\rm loc} \gg 1$.
We comment that 
$\gamma_{\rm loc} \approx \gamma$ is expected
in usual paramagnetic heavy Fermion systems 
where magnetic fluctuations are not prominent
because in such a case the term $\Gamma_{i,M;j,M'}(0;0)$ 
with $i\ne j$,
which represents the inter-site magnetic correlations,
will be small
 \cite{Kontani-susc}.
In fact, the universal KW relation 
$A/\gamma^2 \sim 1\times10^{-5} \mu\Omega$cm[mol$\cdot$K/mJ]$^2$
in many Ce- and U-based HF systems
means the validity of the approximation 
introduced in the present work.
%that the relation $\gamma_{\rm loc} \approx \gamma$ is satisfied
%in many Ce- and U-based HF.

We comment that a similar strong coupling analysis was
performed to derive the Wilson ratio 
($W_{\rm R}=(\chi/\gamma)(2\pi^2k_{\rm B}^2/\mu_{\rm eff}^2)$)
in the present model,
and the relation $W_{\rm R}\approx 1+1/(N-1)$
is derived on condition that the
inter-site magnetic-correlations are weak
 \cite{Kontani-susc}.
This result will be consistent with 
a smaller $W_{\rm R}$
($W_{\rm R}{\raisebox{-0.75ex}[-1.5ex]{$\;\stackrel{<}{\sim}\;$}}1$) 
in YbCu$_{5-x}$Ag$_x$, considering that
$\mu_{\rm eff}$ will be slightly smaller than 4.54$\mu_{\rm B}$ 
(for a free Yb$^{3+}$ ion) due to a small but finite CEF
 \cite{Tsujii2}.
%It is found that both the Pauli and van-Vleck susceptibilities
%are strongly enhanced due to $U$.

Next, we analyze the imaginary part of the self-energy.
Its $T^2$-term within the DMFA is give as
%composed of $g(\e)$'s and
%$\Gamma_{\rm loc}$'s as
 \cite{Yamada-PAM}
\begin{eqnarray}
{\rm Im}\Sigma_M(0) = {\rm Im}\Sigma(0) 
= \frac{\pi(\pi T)^2}{2}(N-1)
 \Gamma_{\rm loc}^2(0,0) \rho^f(0)^3
 \label{eqn:ImS},
\end{eqnarray}
as shown in Fig. \ref{fig:diagram}(d).
Using eq. (\ref{eqn:ImS}),
we derive the expression for $A$
within the DMFA.
According to the Kubo formula, 
the conductivity $\sigma$ is given by
\begin{eqnarray}
\sigma= \frac{e^2}{N_0}\sum_\k \int\frac{d\e}{\pi}
\left(-\frac{\d f}{\d \e}\right)
 |G_\k^c(\e)|^2 \left(\frac{\d \e_\k}{\d k_x}\right)^2
 \label{eqn:sigma} ,
\end{eqnarray}
where $f(\e)=(e^{\e/T}+1)^{-1}$ and
$G_\k^c(\e)= (\e+\mu-\e_\k-V_0^2/(\e+\mu-E_f-\Sigma(\e)))^{-1}$
is the Green function for $c$-electrons.
Note that vertex corrections for currents
are dropped in eq. (\ref{eqn:sigma}),
which is allowed within the DMFA
 \cite{Vollhardt,Kotliar}.

The coefficient $A \equiv \rho/T^2$ is given by
eqs.(\ref{eqn:sigma}) and (\ref{eqn:ImS}).
Assuming the spherical Fermi surface and
using the relation
$(N/2)\rho^f(0)= \rho^c(0) V_0^2/(\mu-E_f-\Sigma(0))^2$
($\rho^c(0)$ being the DOS for $c$-electron per spin),
we obtain that
\begin{eqnarray}
A= {3\pi^7}{k_{\rm F}^{-4}} N(N-1)
 \Gamma_{\rm loc}^2(0,0) \rho^f(0)^4
 \label{eqn:A},
%\rho= {2\pi^2}{3^{-1/3}n^{-4/3}} N\rho^f(0)
% {\rm Im}\Sigma(0)
% \label{eqn:rho}
\end{eqnarray}
where $k_{\rm F}$ is the Fermi momentum.
The number of electrons per unit volume in the present model
is given by $n=k_{\rm F}^3/3\pi^2$.
By assuming the free electron model for the conduction electrons
and reviving $h$ and $k_{\rm B}$,
we obtain the following ``generalized KW relation'' 
in the strong coupling case:
\begin{eqnarray}
A\gamma_{\rm loc}^{-2}
&=& \frac{h}{e^2 k_{\rm B}^2}\cdot
 \frac{9 (3\pi^2)^{-1/3}}{n^{4/3}a^3 N_{\rm A}^2}
 \frac{1}{\frac12 N( N-1)} 
 \nonumber \\
&\approx&  \frac{1\times10^{-5}}{\frac12 N( N-1)} \ \ \
 \mu\Omega{\rm cm[mol}\cdot{\rm K}{\rm /mJ]}^2
 \label{eqn:Agamma},
\end{eqnarray}
where both $\Gamma_{\rm loc}$ and $\rho^f(0)$ are cancelled out.
%In deriving the second line of eq. (\ref{eqn:Agamma}),
%$N_{\rm A}$ is the Avogadro number.
Here, we have used $h/e^2=2.6\times10^4 \ \Omega$,
$k_{\rm B} = 1.38\times10^{-23} \ $JK$^{-1}$,
and assumed that $n^{-1/3} \approx a \approx 1\times10^{-8}$cm
($a$ being the lattice spacing).

According to eq. (\ref{eqn:Agamma}),
$A\gamma_{\rm loc}^{-2} \approx  1\times10^{-5} 
\mu\Omega{\rm cm[mol}\cdot{\rm K}{\rm /mJ]}^2$
for $N=2$ $(J=1/2)$, which corresponds to the 
Kramers doublet ground state case 
due to strong CEF.
On the other hand, 
$A\gamma_{\rm loc}^{-2} \approx  0.36\times10^{-6} 
\mu\Omega{\rm cm[mol}\cdot{\rm K}{\rm /mJ]}^2$
for $N=8$  $(J=7/2)$, which corresponds to 
Yb-based HF systems with weak CEF.
This result is consistent with the experiments
reported in ref.
 \cite{Tsujii}.

\begin{figure}
%\vspace{10mm}
\begin{center}
\epsfig{file=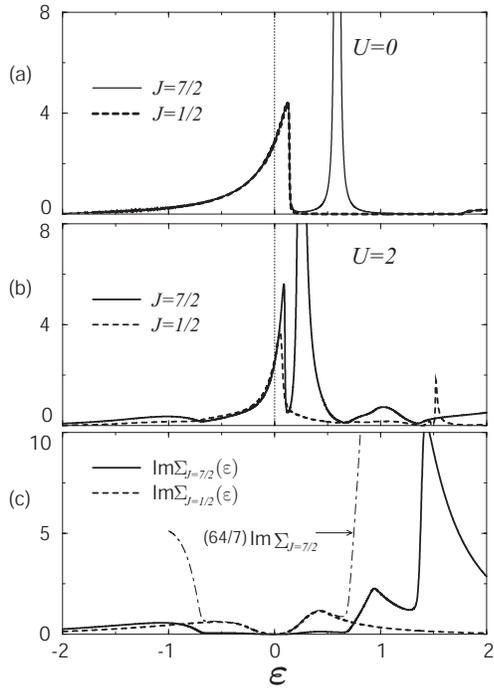,width=8cm} 
\end{center}
%\vspace{10mm}
\caption{
Total $f$-electron DOS ($N\rho^f(\e)$)
for (a) $U=0$ and (b) $U=2$.
A renormalization is recognized in the case of $U=2$.
(c) Im$\Sigma(\e-i\delta)$ given by SOPA ($U=2$).
$\e=0$ corresponds to the Fermi level.
} 
  \label{fig:dos}
\end{figure}

%SOPA
In the next stage,
we study the KW relation
in the weak coupling region ($1/z\simge1$)
using the second-order-perturbation-approximation
(SOPA) with respect to $U$,
both for $J=1/2$ case and $J=7/2$ case
 \cite{Kontani-SOPA}.
In the numerical calculation, 
we use the spherical Brillouin zone
($|\k|\le\pi$) for simplicity of the numerical calculation.
We put $\e_\k= -4+8(\k/\pi)^2$ (the bandwidth being 8),
$E_{\rm f}=-2.5$, $V_0=1.8$ and $n=1.15$ ($n_f=0.8$).
Hereafter, we replace
$E_f$ in eq. (\ref{eqn:Ham}) with  $E_f+ 0.005\e_\k$
for the sake of convenience of numerical calculations.
Figure \ref{fig:dos}
shows the total $f$-electron DOS, $N\rho^f(0)$,
both for $U=0$ and for $U=2$ obtained by the SOPA at zero temperature.
The non-interacting DOS for $J=7/2$
coincides with that for $J=1/2$
except for the sharp peak around $\e= E_f \sim 0.6$
which is given by $G^{f0}(\e)$ 
in eq. (\ref{eqn:green1}).
The bottom (top) of the hybridization gap is 
$\e=0.17$ ($1.84$) for $U=0$.
We note again that $E_{\rm f}>\mu$ in the present calculation
by considering the particle-hole transformation for 
$4f^{13}$-electrons in Yb$^{3+}$ (J=7/2).
In the case of $U=2$,
the DOS around the Fermi level and $E_f$ level
are renormalized within a smaller energy width
due to the $\e$-dependence of Re$\Sigma(\e)$
  \cite{Kontani-SOPA,Saso}.

\begin{figure}
%\vspace{10mm}
\begin{center}
\epsfig{file=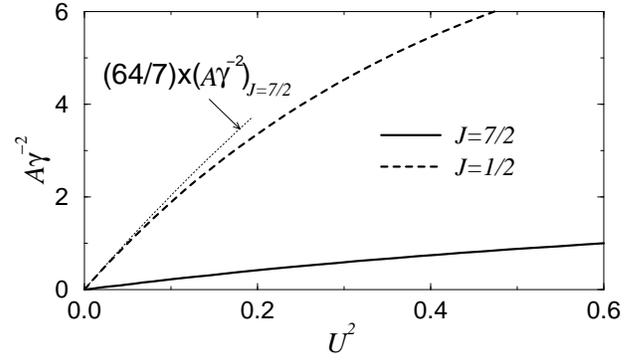,width=8cm}
\end{center}
%\vspace{10mm}
\caption{
$A\gamma^{-2}$ obtained by SOPA as functions of $U^2$.
$A\gamma^{-2}$ for $J=7/2$ is about 10 times smaller
than that for $J=1/2$ in the weak coupling region. 
}
  \label{fig:KW}
\end{figure}

Figure \ref{fig:dos} (c)
shows the imaginary part of the self-energy for $U=2$
obtained by the SOPA,
which is given by
%is given by replacing $\Gamma_{\rm loc}^2$ in eq. (\ref{eqn:ImS})
%
\begin{eqnarray}
{\rm Im}\Sigma(\e-i\delta) &=& \pi U^2
 (N-1)\int_0^\e d\w \int_{-\e+\w}^0 d\w' 
 \nonumber \\
& &\times  \rho^f(\w)\rho^f(\w')\rho^f(\e-\w+\w')
 \label{eqn:rrr}.
\end{eqnarray}
Thus,
Im$\Sigma_{J=7/2}(\e)$ coincides with 
$(7/64)$Im$\Sigma_{J=1/2}(\e)$
for $|\e|\simle |E_f-\mu| \sim 0.6$ within the SOPA,
as shown in Fig. \ref{fig:dos} (c).
However, Im$\Sigma_{J=7/2}(\e)$ takes much larger
value for $|\e|\simge 0.6$ by reflecting
the huge weight of $G^{f0}(\e)$ around $E_f$.
The real part of the self-energy is obtained from eq. (\ref{eqn:rrr})
using the Cauchy integral.
Figure \ref{fig:KW}
shows $A/\gamma^2$ as functions of $U^2$, 
where $A$ and $\gamma$ are obtained by the SOPA.
$[A/\gamma^2]_{J=7/2}=(7/64)[A/\gamma^2]_{J=1/2}$
is realized within $O(U^2)$, which holds approximately 
in the weak coupling region where $U^2\simle0.2$.
We find that $1/z_{J=1/2}=1.25$ and $1/z_{J=7/2}=1.13$
for $U=2$.
Taking the result by the SOPA
as well as eq. (\ref{eqn:Agamma}) 
derived by the strong coupling analysis,
we can naturally estimate that 
the ratio $A/\gamma^2$ for $J=7/2$ is about one order
smaller than that for $J=1/2$
in any intermediate coupling case.
%This is the main message of the present work.
A perturbation calculation up to $U^4$-order
will be useful for a detailed study, which is a future problem.

% SU(N)-PAM
Finally, we study the KW relation
for the following SU(N)-PAM:
\begin{eqnarray}
H &=& \sum_{\k \s}\e_{\k}c_{\k M}^\dagger c_{\k M} 
 + \sum_{\k M} E_{\rm f} f_{\k M}^\dagger f_{\k M}
 \nonumber \\
& &+ V_0 \sum_{\k M} \left( 
  f^\dagger_{\k M}c_{\k M} + {\rm h.c.} \right)
 + \frac{U}{2}  \sum_{i M\neq M'} n_{i M}^f n_{i M'}^f
 \label{eqn:Ham-SUN},
\end{eqnarray}
where $M=J, J-1,\cdots, -J$ and $N=2J+1$.
Although this model has been frequently analyzed 
by slave-boson $1/N$-expansion method, 
%, eq. (\ref{eqn:Ham-SUN}),
it is less realistic than eq. (\ref{eqn:Ham})
in that (i) both $c$- and $f$-bands have $N$-fold degeneracy
and (ii) the $c$-$f$ mixing is allowed only for electrons 
with equal $M$.
Apparently, the Green function is diagonal with respect to $M$.
The local $f$-Green function $g(\e)$ is given by
$G^f(\e)$ in eq. (\ref{eqn:green2}), instead of eq. (\ref{eqn:green1}).

By performing the same analysis within the DMFA,
it is shown that eqs.(\ref{eqn:gamma_loc}) and (\ref{eqn:ImS})
are also valid in SU(N)-PAM in the strong coupling limit.
On the other hand, the conductivity is given by 
eq. (\ref{eqn:sigma}) times $N/2$.
By using the relation
$\rho^f(0)= \rho^c(0) V_0^2/(\mu-E_f-\Sigma(0))^2 $
in SU(N)-PAM, we obtain that
\begin{eqnarray}
A&=& {12\pi^7}{k_{\rm F}^{-4}} \frac{N-1}{N}
 \Gamma_{\rm loc}^2(0,0) \rho^f(0)^4 .
\end{eqnarray}
Considering that $k_{\rm F}=(6\pi^2n/N)^{1/3}$ in SU(N)-PAM,
the generalized KW relation in 
in SU(N)-PAM is given by
\begin{eqnarray}
A\gamma_{\rm loc}^{-2}
 = \frac{1\times10^{-5}}{(N/2)^{5/3}(N-1)} \ \ \
 \mu\Omega{\rm cm[mol}\cdot{\rm K}{\rm /mJ]}^2
 \label{eqn:Agamma_SUN} ,
\end{eqnarray}
which is $(N/2)^{2/3}$ times smaller than eq. (\ref{eqn:Agamma}),
so it gives too small a value for $N=8$;
$A\gamma_{\rm loc}^{-2}\sim 0.14\times10^{-6}$.
%For $N=8$, eq. (\ref{eqn:Agamma_SUN}) gives 
%$0.14\times10^{-6}\mu\Omega{\rm cm[mol}\cdot{\rm K}{\rm /mJ]}^2$.
We note again that SU(N)-PAM is not realistic for larger $N$
in that both $c$- and $f$-band have $N$-fold degeneracy.
Nonetheless, this result suggests that
the degeneracy of the conduction band further reduces the
value of $A\gamma^{-2}$.

%Summary
In summary,
we have studied the KW relation in 
HF systems with orbital degeneracy.
By analyzing the OD-PAM, eq. (\ref{eqn:Ham}),
on the basis of the FL theory,
we have derived a generalized KW relation
in the strong coupling limit, eq. (\ref{eqn:Agamma}).
The obtained result naturally explains the remarkably 
smaller value of $A\gamma^{-2}$ observed 
in various Yb-based orbitally degenerate
HF systems reported in ref.
 \cite{Tsujii}.
A numerical analysis using the SOPA was also presented.
Another generalized KW relation has been derived based on
the SU(N)-PAM, eq. (\ref{eqn:Agamma_SUN}), 
which also tells that $A\gamma^{-2}$ becomes drastically smaller
due to the orbital degeneracy.
However, the SU(N)-PAM may be less realistic than the
OD-PAM for larger $N$.
The present simplified OD-PAM will be enough to 
understand a global aspect of the KW relation in 
Ce and Yb-based HF systems.
It is an important future problem to study the effect 
of small but finite CEF splitting on the value of $A\gamma^{-2}$
by numerical methods.

Finally,
we comment on the pressure dependence of $A$ in CeCu$_2$Ge$_2$
 \cite{Jaccard}:
It suggests that the value of $A\gamma^{-2}$ decreases
suddenly when the ground state degeneracy increases 
(i.e., $\Delta_{\rm CEF}<T_{\rm K}$) for P$>$15GPa.  
This interesting behavior will be explained within the framework of the 
present study.
The {\it generalized K-W relation}
proposed in the present work is confirmed 
in various HF compounds with $N=2\sim8$ 
 \cite{Future}.
Its importance will increase further 
as various new compounds with orbital degeneracy 
are discovered in future.

%Acknowledgment
The author is grateful to
K. Yamada, T. Saso, D. Vollhardt, Y. Yoshimura and N. Tsujii
for useful comments and discussions.

%%%%
% references
%%%%%%%%%%%%%%%%%%%%

%%%%%%%%%%%%%%%%%%%%%%%%%%%%%%%%%%%%%%%%%%%%%%%%%%%%%
%%%%%%%%%%%%%%%%%%%%%%%%%%%%%%%%%%%%%%%%%%%%%%%%%%%%%%
%%%%%%%%%%%%%%%%%%%%%%%%%%%%%%%%%%%%%%%%%%%%%%%%%%%%%
%%%%%%%%%%%%%%%%%%%%%%%%%%%%%%%%%%%%%%%%%%%%%%%%%%%%%%
%%%%%%%%%%%%%%%%%%%%%%%%%%%%%%%%%%%%%%%%%%%%%%%%%%%%%
%%%%%%%%%%%%%%%%%%%%%%%%%%%%%%%%%%%%%%%%%%%%%%%%%%%%%%

\end{multicols}

\end{document}